\pdfoutput = 1
\documentclass[aps,pra,
twocolumn,
showpacs,superscriptaddress,10pt]{revtex4}

\usepackage{graphicx, subfigure}
\usepackage{color}
\usepackage{amsmath}
\usepackage{amsfonts}
\usepackage{bm}
\usepackage{enumerate}
\usepackage{array}
\usepackage[caption=false]{subfig}
\usepackage{tikz}
\usetikzlibrary{decorations.pathmorphing,arrows,automata,calc,positioning}

\newcommand{\ket}[1]{\ensuremath{|#1\rangle}}

\begin{document}

\title{Photo-activated biological processes as quantum measurements}

\author{A. Imamoglu}
\affiliation{Institute for Quantum Electronics, ETH Z\"urich, CH-8093 Z\"urich, Switzerland}
\author{K. B. Whaley}
\affiliation{Berkeley Quantum Information and Computation Center \\
Department of Chemistry, University of California, Berkeley, CA 94720, USA}

\date{\today}

\begin{abstract}

We outline a framework for describing photo-activated biological
reactions as generalized quantum measurements of external fields,
for which the biological system takes on the role of a quantum
meter. By using general arguments regarding the Hamiltonian that
describes the measurement interaction, we identify the cases where
it is essential for a complex chemical or biological system to
exhibit non-equilibrium quantum coherent dynamics in order to
achieve the requisite functionality. We illustrate the analysis by considering measurement of the solar radiation field in photosynthesis and measurement of the
earth's magnetic field in avian magnetoreception.
\end{abstract}


\maketitle

 \section{Introduction}
 \label{sec:intro}

The role of quantum dynamical effects in biological processes has
generated increasing interest in recent years as time-resolved
measurement techniques have allowed probing of dynamics on
ultra-short time scales~\cite{fleming2011quantum}.
Photo-induced processes are particularly
amenable to such studies, e.g., with pulsed lasers. Many key
biological sensing and regulation processes are initiated by
absorption of visible or near infrared light: these include vision,
photosynthesis and the proposed mechanism for magneto-reception.
While the molecular context of these photo-induced biological
processes may be quite different, they share several important
common features.  Most importantly, the excited state dynamics
following what is typically an electronic excitation of a
chromophore molecule within a pigment-protein complex results in
initiation of a sequence of chemical reactions that result in
biological functionality - be it signaling of external stimuli as in
the cases of vision and magneto-reception, or production of energy
rich compounds in the case of photosynthesis.  The relevant
dynamics following the photo-excitation to the excited state occur
in strongly non-equilibrium conditions. An open question that is at
the heart of the burgeoning field of quantum biology is whether
quantum coherence during this non-equilibrium evolution is relevant
in conveying information about the external stimulus to the specific
molecular components that initiate subsequent biological function.

In this Article, we analyze the relevance of quantum coherent
dynamics in the general class of photo-activated biological
processes by embedding the problem in a quantum measurement setting,
where the light-sensitive bio-molecule takes on the role of a
quantum meter that allows the biological organism to acquire
information and/or energy from the external stimuli. We shall first
outline the key aspects of quantum measurement analogy that will be
used in our analysis.  Following this we consider two categories of
biological quantum measurements. In the first category, the external
stimulus for the bio-system consists exclusively of the
non-equilibrium radiation field which ensures optical pumping of the
pigment-protein complex into a metastable state. This category
includes the light harvesting process that initiates photosynthesis,
as well as the photo-initiation of vision.  In the case of
photosynthesis, the requisite biological function is energy storage;
the underlying irreversible dynamical process can then be classified
as an non-referred quantum measurement, since the information gained
about the incident radiation field is per se not directly relevant,
although the measurement event and its output of an electron hole
pair is essential for the biological function. In contrast, the
information gain about the incident light is central to the retinal
photo-isomerization in vision. The key feature of all biological
processes in this category is that the measurement interaction does
not commute with the Hamiltonian describing the eigenstates of an
unperturbed quantum meter. In the second category, which includes
magneto-reception, the measurement interaction by itself does not
lead to coupling of different meter eigenstates prior to optical
excitation, i.e., the measurement interaction commutes with the
Hamiltonian describing the free quantum-meter evolution.
A key conclusion of the present work is that quantum
coherence is essential for biological processes belonging to the
second category whereas its role in the first category is limited to
enhancement of the information/energy extraction rate.

\section{Quantum measurements and photo-activated biological processes}
 \label{sec:measurement}

Quantum mechanics postulates that  the state of any physical system
at time $t$ is described by a density operator $\rho(t)$; the
probabilities associated with the possible outcomes of an arbitrary
measurement carried out on this system is contained in $\rho(t)$.
Quantum mechanics also postulates that the general time evolution of
a physical system is describable as a quantum operation, specified
by a set of Kraus operators, that relate the initial  and final
density operator~\cite{petruccione2002theory}. This formulation
allows us to treat the dynamics of a physical system that is in
constant interaction with other, possibly larger, physical systems,
which we refer to as the environment; typically, we have no control
over the environment degrees of freedom and no possibility to make a
measurement. As a consequence of these uncontrolled interactions
between the system and the environment, the entropy of the system
increases with time, signalling information about the system leaking
into the environment degrees of freedom. This process, which is
termed decoherence, can encompass both dephasing and relaxation
components and is indistinguishable from a measurement carried out
on the physical system, provided that the measurement results are
discarded. The measurement processes are in turn described by a
Positive Operator Valued Measure (POVM), whose elements are directly
linked to the Kraus operators associated with the underlying quantum
operation. When the elements of the POVM are projection operators
onto the eigenstates of the Hermitian operator $\hat{A}$ associated
with an observable $A$, we say that the quantum operation
corresponds to the measurement of A.

Even though a quantum measurement is normally perceived as
interfacing a system with a classical apparatus, it is convenient to
describe the underlying physical process as a quantum mechanical
interaction between the system to be measured and a {\sl quantum
meter}; as a consequence of the interaction described by the
Hamiltonian $\hat{H}_{meas}$, the system and the meter become
correlated in a way that the post-measurement state of the meter
carries information about the system state. The irreversibility of
the measurement process emerges as a consequence of the coupling of
the quantum meter to other physical systems with large number of
degrees of freedom -- the environment of the quantum meter.  This
formulation was first introduced by von Neumann in 1930s~\cite{john1955mathematical}.

Based on this premise, a large class of chemical reactions  or
biological processes can be described as a quantum measurement where
the biological complex of interest assumes the role of a {\sl
quantum meter}. The simplest scenario with which one can describe a
quantum measurement is the one in which the wave-function describing
the quantum meter is in a pure state - which is normally the lowest
energy eigenstate of the meter Hamiltonian $H_{meter}$. This
assumption does not require that the overall system wave-function is
in a pure state - nor does it assume zero temperature; it is
motivated by the fact that to ensure maximal information extraction
from the system that is being measured, it is desirable to have
complete information about the initial state of the meter. That
said, the assumption of an initial pure-state is well justified in a
pigment-protein complex where the initial photo excitation produces
an electronic excitation of one or more chromophore molecules  from
a non-degenerate electronic ground-state. The rotational and vibrational
degrees of freedom of the bio-molecule on the other hand typically start out and remain in a
mixed-state. For the case of a quantum meter measuring a classical
field such as the earth's magnetic field, suitable meter degrees of freedom are electronic spin and these may initially be in a pure state,
even if the nuclear degrees of freedom are not.  We note that such an electron spin quantum meter is currently of interest in other settings, such as a nitrogen-vacancy center used as a quantum meter to measure weak magnetic fields~\cite{maze2008nanoscale,balasubramanian2008nanoscale}.

To proceed with identifying the relevant Hamiltonian describing a
light-induced biological process, we shall consider the quantum
meter as composed of a system of electrons derived from molecules
within a pigment-protein complex, with associated charge (orbital)
and spin degrees of freedom.   We make the Born-Oppenheimer
approximation and treat the vibrational and rotational degrees of
freedom of the protein as constituting an environment for the system
electrons. The nuclear spins of the environment on the other hand
will be treated separately, as they give rise to an effective
quasi-static internal magnetic field that acts on the electronic
system.
In the interaction picture,
the general Hamiltonian is then
expressed as
\begin{equation}
H=H_{meter} + H_{sys} + H_{meas} + H_{m-env}
\label{eq:general_H}
\end{equation}
where $H_{m-env}$ describes the coupling of the quantum meter to its
{\sl environment} and $H_{sys}$ is  the Hamiltonian of the system to
be measured.
We shall be concerned here with measurement of external stimuli for biological systems, in particular, measurement
of an incident radiation field and of the earth's magnetic field.
The coupling between the system and the meter is
described by the measurement Hamiltonian $H_{meas}$. Since we
exclusively deal with chemical processes that are triggered by
light, we write
\begin{equation}
H_{meas} = H_{m-rad} + H_{int},
\label{eq:H_meas}
\end{equation}
where $H_{m-rad}$ is the electric dipole Hamiltonian describing light
absorption/emission by the pigment-protein complex and $H_{int}$ the
interaction Hamiltonian describing the coupling of the meter to the
external stimuli that is not captured by $H_{m-rad}$.

We assume that the broadband optical excitation arising from
$H_{m-rad}$ projects the quantum meter into a superposition of
eigenstates $|\Psi_{m}^{ex}\rangle$ with eigenenergies that are
substantially higher than that of the initial ground state
$|\Psi_m^{g}\rangle$ of the meter. Before the optical excitation,
the meter dynamics is described by $H_{meter} = H_{0}$. The structural
changes induced by the optical excitation are implicit in the post-excitation meter Hamiltonian  $H_{meter}^{(ex)} =
H_{ex}$. We shall also assume that in general $[H_{ex},H_{0}] \neq
0$, implying that a good quantum number for the ground state manifold
need not be a conserved quantity for dynamics in the relevant metastable excited state manifold.

The need to assign two non-commuting Hamiltonians to the ground and
excited manifolds of the electronic system stems from the influence
of the degrees of freedom that are excluded from the description of
the quantum meter. This situation can arise when the electronic
degree of freedom of interest is spin, which is subject to
interactions whose magnitude strongly depend on the orbital degrees
of freedom of the electron.
For example, before charge separation in the excited
state manifold takes place, the dominant interaction between remote electron spins is electron exchange, whereas following charge separation that results in formation of a radical pair, hyperfine
interaction with neighboring nuclear spins could become the leading electron spin interaction term. More generally,
optical excitation typically leads to changes in the electronic or
nuclear degrees of freedom that are not directly interacting with
the external electromagnetic field but are nevertheless indirectly
affected by the optical excitation.
A mean-field treatment of these additional degrees of freedom would
then yield a modified Hamiltonian for the meter, which would be
described by $H_{ex}$, while the pre-optical-excitation Hamiltonian
was $H_0$. This scenario is akin to a quantum quench induced by
absorption of a photon \cite{latta2011quantum}.

The formulation of the photo-activated measurement process in terms
of the general Hamiltonian, Eq.~(\ref{eq:general_H}), allows us to consider two scenarios:

(i) In the first scenario, the system to be measured is the strength
of the incident light field, or equivalently, the number of photons $n_i$ at specific frequencies $\omega_i$ incident upon the meter:
$H_{sys} = \sum_i \hbar \omega_i n_i$.  In this case, $H_{meas} = H_{m-rad}$ and $H_{int} =0$. Since  $[H_{0},H_{meas}] \neq 0$, the
absorption process in general leads to correlations between the
measured system (incident light) and the meter together with its environment (the total pigment-protein complex). Depending on the statistical properties of the incoming light, upon absorption the pigment-protein complex may be in a classical mixture of states, for thermal light, or a coherent superposition of states, for coherent light~\cite{brumer2012molecular}. Due to the reservoir coupling $H_{m-env}$, the excited meter states relax over a time scale $\tau_0$ into a  metastable
eigenstate $|\Psi_m^d(\tau_0)\rangle$ (for notational simplicity we refer to pure states) that constitutes the doorway
state for subsequent chemical signaling.
We claim that, independent of the coherence in its time evolution, the optical excitation of
the meter followed by this non-radiative relaxation into the
metastable doorway state is essentially an optical pumping process.
The efficiency of the optical pumping is relevant since the
metastable states thereby prepared represent the information gained
and encoded by the meter and facilitate the relevant conditional
chemical reaction which constitutes the signaling step controlling
the subsequent biological function.  The interplay between incoherent and quantum coherent evolution
taking place for $t \le \tau_0$ quantitatively determines the
efficiency of the optical pumping. However, the overall measurement
process can nevertheless be efficiently described in terms of rate equations.
The best known examples of
biological processes that can be described using this scenario
are light harvesting and the primary stages of vision.

(ii) In the second scenario, the coupling between the system to be
measured and the quantum meter is described by $H_{int}$, which
satisfies $[H_{0},H_{int}] = 0$.  In this case, both the meter ground
state $|\Psi_m^g\rangle$ and the excited state
$|\Psi_m^d(\tau_0)\rangle$  that could be reached after the action of $H_{m-rad}$
and $H_{m-env}$ are eigenstates of $H_{int}$ and of $H_{meter}^{(0)}= H_{0}$. Clearly, if $H_{0} = H_{ex}$, the meter cannot aquire information about the system. If on the other hand, the Hamiltonian that governs the dynamics in
the  optically excited manifold ($H_{meter} = H_{meter}^{(ex)} = H_{ex}$)
satisfies $ [H_0, H_{ex}] \neq 0$, then the state
$|\Psi_m^d(\tau_0)\rangle$ will be a superposition of the
eigenstates of $H_{ex}$. Subsequent evolution under $H_{ex}$ then
generates non-trivial quantum dynamics that is sensitive to
$H_{int}$ and could allow for an interferometric measurement. To see
that the extraction of information about the system in this case
relies crucially on the preservation of quantum coherence, we note
that an interferometric measurement  projecting the system back into
the eigenstates of $H_{0}$ requires that the coherence time $\tau_c$
satisfies
 \begin{equation}
 \label{eq:interferometric}
    \tau_{c} > 1/||H_{ex}||
 \end{equation}
assuming $||H_{ex}|| > ||H_{int}||$. A sizeable
$H_{int}$ induced coupling, or accumulated relative phase, between
the eigenstates of $H_{ex}$ is obtained provided $||H_{int}||
\tau_{c} \sim 1$. Equivalently, a measurement within the
lifetime of the excited state is possible if the system retains its
quantum coherence on time-scales long compared to the characteristic
time-scales of the final-structure Hamiltonian $H_{ex}$, indicating
that non-equilibrium quantum dynamics in the optically excited
states is an essential feature. This case presents an especially
intriguing situation for measurement of an external magnetic field
by a biological quantum meter consisting of electron spins. We note that even though the nature of the information extraction in
the measurement process is drastically different in scenario (ii)
than in scenario (i), optical pumping also plays a role in
scenario (ii), via the preparation of the metastable excited state
$|\Psi_m^d(\tau_0)\rangle$.

\section{Measurement of incident radiation field}
\label{sec:lightharvesting} First, we consider the  scenario (i)
where the system observable to be measured is the  mean photon
number of the incident radiation field. The system-meter
correlations that emerge in this case may be interpreted by analogy
to a simple three-level system, in which the central dynamical
processes are the optical excitation of the pigment-protein complex
to a set of high energy states and a subsequent fast non-radiative
relaxation to a lower energy metastable state that acts as a
doorway state for subsequent chemical and biological signaling
processes. Without loss of generality, we may consider incoherent
optical excitation from the ground state $\ket{1}$ to a single high
energy state $\ket{3}$ which we introduce to represent the set of
short-lived excited states. The excitation rate is $\Gamma_{31}
\overline n_{31}$, where $\Gamma_{31}$ is the spontaneous emission
rate from $\ket{3}$ to $\ket{1}$ and $\overline n_{31}$ denotes the
steady state mean photon occupancy. Relaxation from state $\ket{3}$
to the lower energy doorway state $\ket{2}$ takes place through a
combination of quantum coherent evolution due to $H_{meter}$ and
coupling to low energy vibrational degrees of freedom of the
molecule ($H_{m-env}$). Postponing the discussion of potential
quantum effects, we describe this relaxation with a non-radiative
decay rate $\Gamma_{32}$. These two rates, together with the
relatively slow decay of $\ket{3}$ back to the ground state or the decay of $\ket{2}$ to
further reaction product states $\ket{X}$, are features common to
each of the photo activated processes that we consider in this
paper.

To the extent that the relaxation of the doorway
state $\ket{2}$ is slow, this level scheme together with their relevant
couplings corresponds to an optical pumping
scheme (see Figure~1 for two specific examples). A steady state rate equation analysis for the
state populations $\rho_{ii} $ shows that in all cases that we
consider, the population of the doorway state $\ket{2}$ is
proportional to the number density of the incident radiation field
$\overline n_{31}$. In our measurement-based description of photo
activated processes in biology, optical pumping from the ground
state $\ket{1}$ to the metastable excited state $\ket{2}$ is thus
equivalent to a measurement of the incident solar radiation field
observable $\overline n_{31}$, or equivalently its temperature.

\subsection{Light harvesting complexes in photosynthesis}
\label{subsec:Radiationfield_photosyn} To illustrate this
measurement of incident light intensity via optical pumping in a
biological setting, we first consider the light harvesting step in
photosynthesis.  Here the quantum meter is the chromophore component
of a pigment-protein complex known as the light harvesting complex
(LHC) or the `antenna complex', whose role is to transfer the energy
from absorbed photons to the reaction center where  subsequent
separation of the electron-hole pair occurs. The system to be
measured is the out-of-equilibrium source of light -- typically
sunlight filtered by the earth's atmosphere, although some bacteria
living deep below the ocean near hydrothermal vents employ black
body radiation from these vents~\cite{beatty2005obligately}. The information extraction in this case is
accompanied by energy storage in the quantum meter. This is brought
about by subsequent steps transferring the electronic energy from
the antenna complex to the reaction center where charge separation
occurs, generating electrons that initiate chemical reactions
leading to energy rich products. The average population of the
photo-generated electrons in the reaction center can then be
considered to represent the encoding of information gained from a
measurement of the mean number of absorbed photons (typically
visible or near infrared).

\newcommand{\heightStateThree}{5}
\newcommand{\heightStateTwo}{3.75}
\newcommand{\heightStateT}{2.5}
\newcommand{\heightStateX}{0.8}

\begin{figure}
\subfigure[\,Optical pumping in light harvesting]{
  \centerline{
    \begin{tikzpicture}[
      scale=.9,
      level/.style={very thick},
      virtual/.style={very thick,densely dashed},
      classical/.style={very thick,double,<->,shorten >=4pt,shorten <=4pt,>=stealth},
      transition/.style={very thick, black,->,>=stealth',shorten >=1pt,color=red, text=black},
      radiative/.style={transition,color=blue, text=black,decorate,decoration={snake,amplitude=1.0}},
    ]
    \draw[level] (3,0) -- (0,0) node[left] {\ket{1}};
    \draw[level] (5,\heightStateThree) -- (2,\heightStateThree) node[left] {\ket{3}};
    \draw[level] (4,\heightStateT) -- (5.5,\heightStateT) node[right] {\ket{2}};
    \draw[transition] (0.5,0) -- (2.5,\heightStateThree) node[midway,left] {$\bar{n}_{31} \Gamma_{31}$};
    \draw[radiative]  (3,\heightStateThree) -- (1,0) node[midway,right] {$\Gamma_{31}$};
    \draw[radiative] (4,\heightStateThree) -- (4.75,\heightStateT+.05) node[midway,left] {$\Gamma_{32}$};
     \draw[radiative] (4.75,\heightStateT) -- (5.65,0.5) node[midway,left] {$\Gamma_{2x}$};
    \draw[radiative] (4.4,\heightStateT) -- (2.5,0) node[midway,right] {$\Gamma_{21}$};
    \end{tikzpicture}
  }
}
\par\bigskip
\subfigure[\,Optical pumping in vision]{
\centerline{
  \begin{tikzpicture}[
    scale=.9,
    level/.style={very thick},
    virtual/.style={very thick,densely dashed},
    classical/.style={very thick,double,<->,shorten >=4pt,shorten <=4pt,>=stealth},
    transition/.style={very thick, black,->,>=stealth',shorten >=1pt,color=red, text=black},
    radiative/.style={transition,color=blue, text=black,decorate,decoration={snake,amplitude=1.0}},
  ]
  \draw[level] (3,0) -- (0,0) node[left] {\ket{1}$\equiv$\ket{c}};
  \draw[level] (5,\heightStateThree) -- (2,\heightStateThree) node[left] {\ket{3}};
  \draw[level] (4,\heightStateTwo) -- (5.5,\heightStateTwo) node[right] {\ket{2}};
  \draw[level] (5,\heightStateT) -- (6.6,\heightStateT) node[right] {\ket{t}};
  \draw[level] (6,\heightStateX) -- (7.5,\heightStateX) node[right] {\ket{X}};
  \draw[transition] (0.5,0) -- (2.5,\heightStateThree) node[midway,left] {$\bar{n}_{31} \Gamma_{31}$};
  \draw[radiative]  (3,\heightStateThree) -- (1,0) node[midway,right] {$\Gamma_{31}$};
  \draw[radiative] (4,\heightStateThree) -- (4.75,\heightStateTwo) node[midway,right] {$\Gamma_{32}$};
  \draw[radiative] (4.25,\heightStateTwo) -- (2.5,0) node[midway,right] {$\Gamma_{2c}$};
  \draw[radiative] (4.75,\heightStateTwo) -- (5.75,\heightStateT+0.005) node[midway,right] {$\Gamma_{2t}$};
  \draw[radiative] (5.75,\heightStateT) -- (6.8,\heightStateX+0.004) node[midway,right] {$\Gamma_{tX}$};
  \end{tikzpicture}
}
}
\caption{  
\small{(a) Schematic of optical pumping for light harvesting. The
rate constants for this optical pumping scheme are as follows:
$\Gamma_{31}, \Gamma_{21} \sim (ns)^{-1}, \Gamma_{2X} \sim (1 - 4~
ps)^{-1}, \Gamma_{32} \sim (100-800~ps)^{-1}$~\cite{engel2007evidence,bennett2013structure}.
\\\hspace{\textwidth}    
(b) Optical pumping
scheme corresponding to photon-absorption induced dynamics of
rhodopsin. The magnitude of the key decay rates are $\Gamma_{32}
\sim (80~fs)^{-1}$~\cite{polli2010conical}, $\Gamma_{2t} \sim (140~fs)^{-1}, \Gamma_{2c} \sim
(280~ps)^{-1}$~\cite{note4}, while the remaining rate constants are $\Gamma_{31} \sim (20~ps)^{-1}$~\cite{logunov2001relaxation} and $\Gamma_{tX} \sim (1~ps)^{-1}$~\cite{mccamant2011re}.}} 
\label{fig:opticalpumping_lh}
\end{figure}
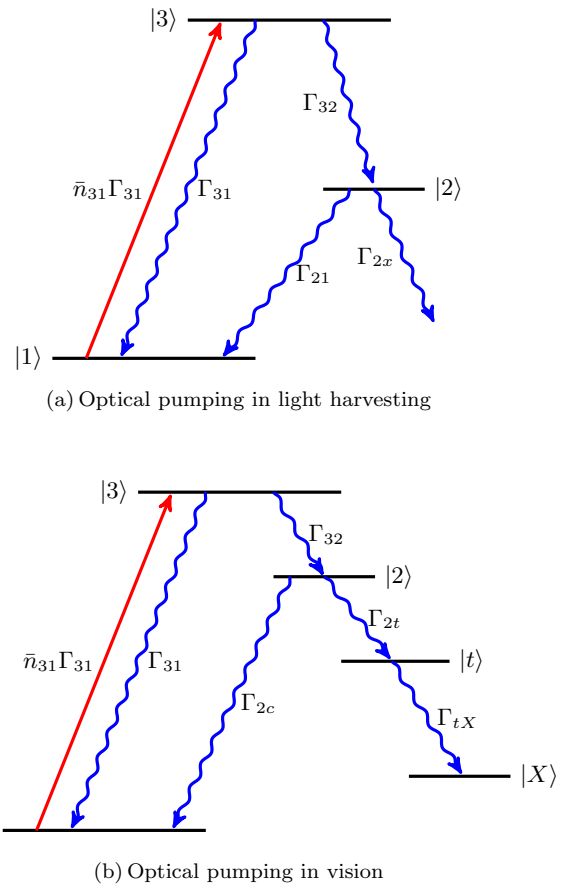

Figure \ref{fig:opticalpumping_lh}a shows the basic scheme for
interpretation of light  harvesting as optical pumping of the
doorway state leading to charge separation. In the notation of
scenario (i), the LHC meter is initially in state $\ket{1}$ in which
the pigments are in their electronic ground states.  The meter is
excited into a metastable state $\ket{3}$ corresponding to a
superposition of excitonic eigenstates of the
pigment subsystem by absorption of broadband photons. The meter then
exhibits  complex non-equilibrium quantum dynamics during which this
initial superposition relaxes into the doorway state $\ket{2}$. This
may may then undergo radiative decay back to the ground state
(fluorescence) with rate $\Gamma_{21}$, or non-radiative
transformation to further products $X$, with rate $\Gamma_{2X}$. The
relaxation process from $\ket{3}$ to $\ket{2}$ is known to be
characterized by a remarkable near-unity quantum efficiency; within
our optical pumping model, this implies that the short time dynamics
must yield complete transfer of $\rho_{33}$ to $\rho_{22}$.  Unlike
most optical pumping schemes in atomic physics though, the ambient
conditions relevant for biological processes ensure that $\rho_{22}
\ll \rho_{11}$.

The key question of interest for us is the role of  quantum coherence during this measurement process.
In particular, to what extent does the quantum nature of the meter
(LHC) play an enabling role in the measurement of the incident light
field? The non-equilibrium energy transfer dynamics of light
harvesting have been extensively studied in recent
years~\cite{cheng2009dynamics,ishizaki2010quantum,scholes2011lessons} and a full description was shown to require
simulation of  complex open quantum system dynamics with both
coherent and incoherent components. We note
however, that the interplay between dipole-dipole interaction
($H_{ex}$) mediated inter-chromophore exciton hopping and the
coupling to vibrational degrees of freedom ($H_{m-env}$) ensure the
energy transfer from the initially excited chromophore state
$\ket{3}$ to the reaction-center state $\ket{2}$, irrespective of
the relative magnitude of the coherence time $\tau_c$ to
characteristic quantum coherent evolution timescale
$||H_{ex}||^{-1}$. In fact, theoretical studies have shown that
preserving quantum coherence, or equivalently, reducing the effects
of dephasing and dissipation in the light harvesting produces a
relatively small quantitative change in efficiency rather than an
on/off switch of functionality~\cite{note1}.

\subsection{Photo-activated isomerization in vision}
\label{subset:Radiationfield_vision} Photo-activated  isomerization
constitutes another class of biological processes that may be
understood in terms of optical pumping realizing a measurement of
incident mean visible photon number ($\overline n_{31}$) by a
biological quantum meter. Light activated isomerization reactions
play an important role in control and switching of biological
function in a broad range of organisms, including vision in animals
and photosynthesis in halo bacteria.
A prime
example of such a photo-activated isomerization based quantum meter
is the rhodopsin pigment-protein complex which plays the key role in
the primary steps of animal vision~\cite{schoenlein1991first}.

Figure~1b shows the energy level diagram of the relevant states
involved in the photo activation of vision by retinal in rhodopsin
protein. In the ground state $\ket{1} \equiv \ket{c}$ of rhodopsin,
the retinal pigment is in the cis-conformation. Experimental studies
have shown that the transformation from initial Franck-Condon
photo-product $\ket{3}$ reached by photon absorption, to the
metastable all-trans isomer of retinal proceeds via a conical
intersection, which is reached within ~$\sim$ 80 fs\cite{polli2010conical}. The
system returns to the ground electronic state nuclear potential
energy surface $V_{S_0}$ by traversing the conical intersection,
arriving in the transitory state, labeled $\ket{2}$ in Fig~1b. This state then undergoes rapid bifurcation, with
approximately 65\% chance of undergoing nuclear dynamics
transforming it into the trans isomer $\ket{t}$ which constitutes
the doorway state signalling photon absorption; the remaining 35\%
returns to the cis-isomer of the electronic ground state,
$\ket{1}$. The overall transformation of $\ket{3}$ to $\ket{t}$
takes place in $\sim$ 200 fs. Combined with a $50\%$
absorption probability for a single photon by an ensemble of
rhodopsin molecules contained within a rod cell~\cite{rieke1998single},
the overall process results in a remarkable $\sim 30\%$ probability
of detection of single photons. While the electronic and vibrational
degrees of freedom are expected to be entangled around the conical intersection~\cite{tully2012perspective}, it is not clear what role, if any,  the underlying quantum correlations play for the creation of the signalling trans state.


\section{Magnetoreception as a quantum measurement of a classical field}
\label{sec:magneto}

Magnetoreception refers to the ability of living organisms to detect
the magnitude and/or orientation of the earth's magnetic field.
Typically found in migrating species, it has been most widely
studied in birds which are capable of navigating distances of
thousands of kilometers~\cite{wiltschko2012global,mouritsen2012magnetic}. This is quite
remarkable, given that the earth's magnetic field is very weak
($\sim 50 \mu$T) and the Zeeman interaction energy of a molecule with
such a field is typically more than 6 orders of magnitude smaller
than $k_B$T. Several biophysical mechanisms have been proposed to
rationalize this remarkable ability~\cite{ritz2000model,kirschvink2001magnetite,kirschvink2014sensory}. One proposed mechanism, the radical pair hypothesis, is
equally remarkable in that it requires maintenance of coherent
quantum spin dynamics over nm distances on time scales (well)
exceeding $10$~ns~\cite{rodgers2009chemical}. While there is so far no unambiguous evidence for
this mechanism {\it in vivo}, there is circumstantial evidence that
it contributes at least partially to avian magnetoreception in some
species~\cite{maeda2012magnetically,niessner2013magnetoreception}, as well as mounting evidence of
feasibility from {\it in vitro} studies with biomimetic molecular
model systems~\cite{maeda2008chemical}.  It thus presents an intriguing and
dramatic instance of non-equilibrium quantum dynamics that may be
essential for biological function.

The molecular basis of the radical pair mechanism is described in a
number of review
articles~\cite{rodgers2009chemical,ritz2011quantum,mouritsen2012magnetic}.
We note that magnetoreception is widely accepted to be
photo-activated, allowing it to be mapped directly into the general
framework for photo-induced biological processes.   Here we present
an analysis of the proposed mechanism within the formalism for
measurement of an external field by a biological quantum meter
described above.  We shall illustrate this with specific reference
to the cryptochrome protein, a photoreceptor which is the leading
candidate for hosting the radical pair in the retina of birds.  This
protein contains a co-factor, FAD, which absorbs incident light
centered around $450$~nm to form an excited singlet state FAD$^*$.
The unstable FAD$^*$ triggers a rapid charge transfer across a chain
of three tryptophan amino acids, leading to the formation of a
radical pair state [FAD$^{\bullet-}$ + TrpH$^{\bullet+}$] in which
the electron spins are located on spatially separated and distinct
molecules. The total electron spin is conserved during this fast
electron transfer, which takes place on a ps time-scale $\tau_0$.

Mapping this onto our measurement scenario (ii), we identify the
electrons of the radical pair as the quantum meter, characterized by
Hamiltonians $H_0$ in the ground state and $H_{ex}$ in the excited state. For $r < 1$~nm, the dominant term in
$H_{ex}$ is well approximated by the exchange interaction
\begin{equation}
H_{ex} \simeq  J(r) \vec{S}_1 \cdot \vec{S}_2 \simeq H_0
\label{eq:meter1}
\end{equation}
where $J(r)$ depends
exponentially on the inter-electron separation $r$. The dominant contribution to
$H_{ex}$  for $r > 1$~nm on the other hand, is given by the anisotropic hyperfine interaction of the two
electron spins with the proximal nuclear spins in their local
environments:
\begin{equation}
H_{ex} \simeq  \sum_{i_1,k}  A_{i_1,k}
S_{1,k}  I_{i_1,k} + \sum_{i_2,k}  A_{i_2,k}
S_{2,k}  I_{i_2,k}. \label{eq:hyperfine}
\end{equation}
Here $I_{i_1}, I_{i_2}$ denote the nuclear spins with non-negligible
coupling to the spin of the unpaired electrons localized at
FAD ($\vec{S}_1$) and tryptophan ($\vec{S}_2$), respectively. ${\bf A}_{i,k}$
is the corresponding hyperfine coupling constant along
$\hat{k}$ ($k = x,y,z$).
The system to be measured here is the earth's magnetic field: this classical field $B_{ext}$ appears in the system-meter interaction term $H_{int}$ of $H_{meas}$:
\begin{equation}
H_{int} = g_e\mu_B \vec{B}_{ext} \cdot (\vec{S}_1 +  \vec{S}_2),
\label{eq:int}
\end{equation}
where we assume that the g-factor of the electron $g_e$ is
independent of its location within the pigment-protein complex
\cite{note2}.
We also discard the much smaller coupling of nuclear
spins to $\vec{B}_{ext}$. Since measurement of the earth's magnetic
field is initiated by sunlight, $H_{meas}$ must also include
$H_{m-rad}$, which is given here by the electric dipole Hamiltonian
describing sun light absorption by the FAD chromophore. The environmental Hamiltonian
$H_{m-env}$ is given by the interactions of the radical pair
electrons with the vibrations and rotations of the cryptochrome
protein. In fact, the preparation of the cryptochrome in the
long-lived [FAD$^{\bullet-}$ + TrpH$^{\bullet+}$] singlet state (state $|2\rangle$
in Fig.~1a) is accomplished by an optical pumping process that is
based on $H_{m-rad}$ and $H_{m-env}$.

We emphasize that since the meter
starts out in a singlet state and since $[H_0,H_{int}]=0$, there are no system-meter correlations before
optical excitation. After the optical pumping process prepares the
pigment-protein complex in the metastable radical-pair [FAD$^{\bullet-}$ +
TrpH$^{\bullet+}$] with an inter-electron distance of $r \sim 1.5$~nm, the
relevant meter Hamiltonian becomes $H_{ex}$, given by
Eq.~(\ref{eq:hyperfine}). Since the timescale for completion of
optical pumping, i.e., the optical excitation followed by radical
pair formation, is much shorter than $||H_{ex}||^{-1}$, the radical
pair is initially in the singlet state, $^S$[FAD$^{\bullet-}$ + TrpH$^{\bullet+}$].
However, since the strength and the anisotropy of the hyperfine
interactions in the FAD and tryptophan molecules are different, the total electron spin is no longer
conserved. Consequently the singlet state is not an eigenstate of
$H_{ex}$ and the hyperfine interactions give rise to dynamic
inter-conversion of singlet and triplet radical states $^{S}$[FAD$^{\bullet-}$ + TrpH$^{\bullet+}$]$\longleftrightarrow^{T}$[FAD$^{\bullet-}$ +
TrpH$^{\bullet}$].

This inter-conversion requires a full quantum
mechanical description over the time scale of the spin coherence of
the radical pair.  Since $[H_{ex},H_{int}] \neq 0$, the weak magnetic field of the
earth $B_{ext}$ modifies the coherent singlet-triplet inter-conversion
provided that its orientation is not parallel to that of the hyperfine (difference) field~\cite{note3}. Since the singlet and
triplet states possess different reaction pathways, this modulation
can cause changes in the populations of the resulting products. In
cryptochrome, both singlet and triplet states can convert to a
long-lived  ($\ge 100 \mu$s) protonated state
[FADH$^{\bullet}$ + Trp$^{\bullet}$], while only the singlet state can undergo
relaxation by back electron transfer to the initial (singlet) ground
state FAD + TrpH~\cite{maeda2012magnetically}.
This singlet relaxation occurs on a time-scale of
$\tau_p \ge 1 \mu s$, so that modulations on shorter time scales can
cause observable changes in the combined singlet and triplet
population that is converted into the long-lived protonated state.
Consequently a detection of the changes in the protonated
radical pair singlet and triplet states or products of subsequent
chemical reactions, constitutes a
measurement of $\vec{B}_{ext}$. The essential feature of this
radical-pair based magnetoreception is thus the dependence of the
long-time-scale ($> 1\mu s$) protonated state population on the relative
orientation of the earth's magnetic field $\vec{B}_{ext}$ with respect to the
radical pair axis.

%

To elucidate the essential role played by quantum coherence in
magneto-reception, we may consider the well-known simplified problem
of two electron spins, (e.g., one at FAD (F) and the other at
Tryptophan (T)), with only one of these electron spins (e.g., F)
interacting with a single nuclear spin $I^F =1/2$. $H_{ex}$ is then given
by the anisotropic hyperfine interaction at site F:
\begin{equation}
H_{ex} =  A_{z}
S_z^{F} \cdot I_z^{F} \label{eq:hyperfinesimple}
\end{equation}
Even though  the nuclear spin is in a completely mixed state, its state remains unchanged during the time scale over which the electron spin evolves; we may therefore assume that it is initially oriented along z, i.e.,
$|\Uparrow\rangle_F$, without loss of generality. The energy levels for this simplified scheme are shown in Figure 2a.

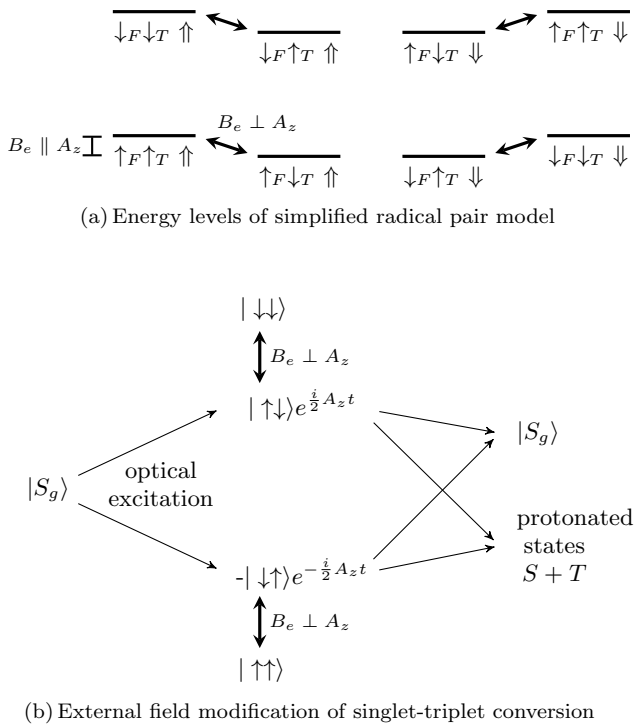
\begin{figure}
\subfigure[\,Energy levels of simplified radical pair model]{
  \centerline{
    \begin{tikzpicture}[
      scale=.55,    
      level/.style={very thick},
      B_transition/.style={very thick,<->,shorten >=4pt,shorten <=4pt,>=stealth},
      delta/.style={thick,|-|},
    ]
    \draw[level] (0.5,1) -- (2.5,1) node[midway, below] {$\uparrow_F\uparrow_T \ \Uparrow$};
    \draw[level] (4,0.5) -- (6,0.5) node[midway, below] {$\uparrow_F\downarrow_T \ \Uparrow$};
    \draw[level] (7.5,0.5) -- (9.5,0.5) node[midway, below] {$\downarrow_F\uparrow_T \ \Downarrow$};
    \draw[level] (11,1) -- (13,1) node[midway, below] {$\downarrow_F\downarrow_T \ \Downarrow$};
    \draw[level] (0.5,4) -- (2.5,4) node[midway, below] {$\downarrow_F\downarrow_T \ \Uparrow$};
    \draw[level] (4,3.5) -- (6,3.5) node[midway, below] {$\downarrow_F\uparrow_T \ \Uparrow$};
    \draw[level] (7.5,3.5) -- (9.5,3.5) node[midway, below] {$\uparrow_F\downarrow_T \ \Downarrow$};
    \draw[level] (11,4) -- (13,4) node[midway, below] {$\uparrow_F\uparrow_T \ \Downarrow$};
    \draw[B_transition] (2.5,1) -- (4,0.5) node[midway,left] {};
    \draw[B_transition] (2.5,4) -- (4,3.5) node[midway,left] {};
    \draw[B_transition] (9.5,0.5) -- (11,1) node[midway,left] {};
    \draw[B_transition] (9.5,3.5) -- (11,4) node[midway,left] {};
    \draw[delta] (0,0.5) -- (0,1) node[midway,left] {\scriptsize$B_e \parallel A_z$};
    \node (Sleft) at (4.,1.3) {\scriptsize$B_e \perp A_z$};
    \end{tikzpicture}
  }
}
\par\bigskip
\subfigure[\,External field modification of singlet-triplet conversion]{
  \begin{tikzpicture}[
  scale=.75,    
  >=stealth',shorten >=2pt,node distance = 2.2cm,
  B_transition/.style={very thick,<->,shorten >=4pt,shorten <=4pt,>=stealth},
  transitionLeft/.style={->,shorten >=5pt},
  transitionRight/.style={->,shorten <=8pt, shorten >=-1.5pt}
  ]
  \node (Sleft) at (0,3.5) {\ket{S_g}};
  \node [label={right:{\ket{S_g}}}] (Sright) at (8,4.5) {};
  \node [label={[text centered,text width=1cm]right:protonated\\states\\$S+T$}] (ProStates) at (8,2.5) {};
  \node (SingletTop) at (4.5,5) {\ket{\uparrow\downarrow}$e^{\frac{i}{2}A_z t}$};
  \node (TripletTop) at (3.82,6.7) {\ket{\downarrow\downarrow}};
  \node (SingletBot) at (4.5,2) {-\ket{\downarrow\uparrow}$e^{-\frac{i}{2}A_z t}$};
  \node (TripletBot) at (3.78,0.3) {\ket{\uparrow\uparrow}};
  \node[text centered,text width=2cm] (leftText) at (2,3.6) {optical excitation};

  \draw[transitionLeft] (Sleft) -- (3.2,5) node[] {};
  \draw[transitionLeft] (Sleft) -- (3.2,2) node[] {};
  \draw[transitionRight] (5.5,2) -- (Sright) node[] {};
  \draw[transitionRight] (5.5,4.95) -- (Sright) node[] {};
  \draw[transitionRight] (5.5,2) -- (ProStates) node[] {};
  \draw[transitionRight] (5.5,4.95) -- (ProStates) node[] {};
  \draw[B_transition] (3.75,5.2) -- (3.75,6.5) node[midway,right] {\scriptsize$B_e \perp A_z$};
  \draw[B_transition] (3.75,1.8) -- (3.75,0.5) node[midway,right] {\scriptsize$B_e \perp A_z$};
  \end{tikzpicture}
}
\caption{  
\small{
(a) Schematic of energy levels relevant to magnetoreception in cryptochrome, within the simplified radical pair model of two electron spins, $S^F$ at FAD and $S^T$ at Tryptophan (single up/down arrows) with a single nuclear spin $I^F$ (double up/down arrow) interacting with the FAD electron spin.
With an anisotropic hyperfine tensor $A \equiv A_z$, a weak external field $B_e \parallel \hat{z}$, just shifts the energy levels (vertical bars), while $B_e \perp \hat{z}$ induces transitions between spin levels with different $z$-projection of $S=S^F + S^T$ (double headed arrows). For $A_z \gg B_e$, the resulting eight total spin levels are divided into two groups separated by a gap of order $A_z$.
\\\hspace{\textwidth}    
(b) Schematic of the resulting external magnetic field modified singlet-triplet conversion in optically excited cryptochrome for $B_{e} \perp \hat{z}$. Upon optical excitation and the subsequent fast relaxation leading to radical pair formation, the protein-chromophore complex is prepared in a coherent superposition of its eigenstates $\ket{\uparrow,\downarrow}$ and $\ket{\downarrow,\uparrow}$. While the hyperfine interaction modifies the relative phase accumulated by $\ket{\uparrow,\downarrow}$ and $\ket{\downarrow,\uparrow}$, $B_{e} \perp A_z$ leads to coherent excitation of the the other two triplet states $\ket{\uparrow,\uparrow}$ and $\ket{\downarrow,\downarrow}$, thereby modifying the probability that the molecule ends up in the protonated state.
}}
\label{fig:opticalpumping_lh}
\end{figure}
Starting out in the singlet state
of the electrons at sites F and T at time $t=0$, the wave-function of the
coupled electron-nuclear system for $B_{ext} =0$ is
\begin{equation} |\Psi(t)\rangle =   \frac{1}{\sqrt{2}} |\uparrow,\Uparrow\rangle_F \otimes |\downarrow\rangle_T e^{-i A_z t}
- \frac{1}{\sqrt{2}} |\downarrow,\Uparrow\rangle_F \otimes
|\uparrow\rangle_T  . \label{eq:wavefunc}
\end{equation}
If $B_{ext} \parallel \hat{z}$, then the initial singlet state can
couple only to a single triplet state, $\ket{T_0}$: assuming equal slow
decay of the singlet and triplet states into distinct chemical
products, this will yield a long-time singlet ($\ket{S}$) yield of $50 \%$.
However, when $B_{ext} \perp \hat{z}$ and in the low field regime
where $B_{ext} \ll A_z$, the Zeeman interaction only influences the
electron at site T, which is not coupled to a nuclear spin. In this case, all three triplet states are populated
with a long-time singlet yield of $25 \%$, provided $A_z \gg B_{ext}
\sim \tau_c^{-1}$. In the opposite limit of $A_z \ll \tau_c^{-1}$ on
the other hand, the singlet-triplet transition probability vanishes to
lowest order. We can therefore conclude that preservation of quantum
coherence over the dynamical timescales associated with $H_{ex}$ is
essential for magneto-reception. Figure 2b illustrates the relevant energy level diagram for $B_{ext} \ll A_z$: the electron spin at site T precesses around the external field, leading to excitation of all $3$ triplet states whenever $B_{ext}$ is not parallel to $\hat{z}$.  The analogy with a simple interferometer is imperfect since all four two-electron spin states are in general coupled by the dynamics (see also~\cite{cai2013chemical}).

To quantify the role of quantum coherence in this simple model, we have carried out a calculation assuming $A_z = 1000~\mu$T and compared the long-time triplet yields for the cases where the external field ($B_{ext} = 50~\mu$T) is parallel or perpendicular to $z$.  We have assumed that the relaxation time back to the initial ground state as well as to the long-lived protonated states is $1~\mu$s. We find that the difference in the triplet yield between the two configurations increases by a factor of $10^8$ when the electron spin coherence time is increased from $\tau_c = 0.1 A_z^{-1} = 1$~ns to $\tau_c = 10 A_z^{-1} = 100$~ns. The strongly nonlinear increase in the sensitivity confirms the essential role played by the quantum coherence.

We emphasize that this description is overly
simplistic, since it neglects the presence of multiple nuclear spins
which reduce the overall directional sensitivity of cryptochrome. In
principle, $H_{ex}$ should also include contributions from exchange
and magnetic-dipole interaction between the separated electrons; we
neglect these contributions here for simplicity, noting however that
exchange interactions could reduce the sensitivity of the
pigment-protein complex to $B_{ext}$. The dipole-dipole interactions
on the other hand could facilitate magneto-reception without the
need for hyperfine coupling due to their inherent anisotropic
nature. Indeed, using numerical calculations assuming isotropic hyperfine interaction ($A = A_x = A_y = A_z$) and strong dipolar interaction with strength $V_{dip} = A = 1000~\mu$T in the same simplistic model, we find that the difference in the triplet yield is comparable to that of the anisotropic hyperfine case discussed earlier for $\tau_c = 10 A_z^{-1} = 100$~ns.

The above analysis of avian reception presents a picture of an array
of quantum meters located in the retina of the bird, each of which
measures the magnitude and orientation of the magnetic field
relative to its own orientation and produces a classical signal in
the form of a chemical population derived from the integrated time
dependence of the protonated radical pair population. One of the underlying assumptions in radical-pair-based magnetoreception is that the bird's brain undertakes processing and integration of all such classical signals
deriving from an array of quantum meters~\cite{wu2012neural}: it thus generates visual
modulation patterns via chemical signaling of the intrinsically
quantum protonated state yield. It is these variations in the modulation
patterns that yield the desired magnetic field information~\cite{ritz2011quantum}.  An
interesting aspect of this biological quantum measurement is that it
is continuous in time, with the cumulative
protonated-state population providing the calibration for the
classical field.

\section{Conclusion}
\label{sec:conclusion}
In this Article, we have developed a general
formalism for describing photo-activated biological processes as a
quantum measurement, where a protein-pigment complex takes on the
role of a quantum meter. We argued that this formulation allowed us
to identify the conditions under which preservation of quantum
coherence and the associated non-equilibrium quantum dynamics
becomes essential for the biological function. Description of the
initial step of the measurement where the protein is prepared in a
doorway state as an optical pumping process enabled us to highlight
the common features of the photo-activated processes.

We have argued that the preservation of quantum coherence during the
time-scale in which  the metastable doorway state
$|\Psi_m^d(\tau_0)\rangle$ is formed is not essential for the
primary biological function when $[H_{0},H_{meas}] \neq 0$;  the
presence or absence of quantum coherence during this time window
only leads to modest quantitative improvements in the efficiency of
the energy storage in light harvesting. Nevertheless, we emphasize
that there may be scenarios in which a small quantitative increase
in efficiency could provide a major advantage to the organism. There
is also the possibility that presence of quantum coherence plays an
important role in imposition of unidirectional energy flow, quantum
ratcheting of energy over uphill steps and enabling long range
transport~\cite{hoyer2012spatial}.

The relevance of the formulation we present here depends  strongly
on the identification of further biological processes for which the
preservation of quantum coherence and the ensuing non-equilibrium
quantum dynamics is essential, i.e. $[H_{0},H_{int}] = 0$ and
$[H_{ex},H_{int}] \neq 0$. At present, the proposed mechanism for
magnetoreception is the only candidate system that is in this class.
On the other hand, we note that optically induced radical pairs are
also sensitive to weak electric fields;  this sensitivity is due to
the different electric dipole moments of the singlet and triplet
states, which in turn results in an external electric field
dependent relative phase between the two spin states. As a
consequence, the singlet-triplet oscillations in the excited state
can be altered by external electric fields. In fact, it is well
known that fluctuating electric fields can lead to dephasing of
singlet-triplet coherence~\cite{PhysRevLett.109.107401}. Taken
alone, such dephasing is equivalent to a non-referred quantum
measurement of the electric field. However, in the context of an
optically induced radical pair in a biological setting with chemical
reaction products dependent on the singlet-triplet dynamics, the
underlying electric field sensitivity of the singlet-triplet energy
difference may facilitate measurement of a local electric field.

\section{Acknowledgements}
\label{sec:acknowledgements} We are grateful to the
Wissenschaftskolleg zu Berlin for unparalleled support  during our
tenure as Fellows in 2012-2013. We have both learned a great deal
from Peter Hore, whose insightful comments were crucial to the
analysis presented in this manuscript.  We thank Donghyun Lee for
assistance with the figures. KBW also thanks the Kavli Institute for
Energy Nanosciences and DARPA for support.


\begin{thebibliography}{37}
\bibitem[{not({\natexlab{b}})}]{note1}
\bibinfo{note}{If dipole-dipole interaction mediated exciton hopping had been
  predominantly interrrupted by pure dephasing processes, there would have been
  a Zeno-type reduction in the exciton transfer rate where the transfer
  efficiency had scaled linearly with the coherence time.}
\bibitem[{not({\natexlab{c}})}]{note2}
\bibinfo{note}{In reality, the g-factor of the electron does change upon its
  transfer from tryptophan to FAD. The resulting gradient of spin-splitting
  however is negligible in the low-field limit of interest.}
\bibitem[{not({\natexlab{d}})}]{note3}
\bibinfo{note}{This simple picture is valid only if the net hyperfine field
  axis remains unchanged for various nuclear spin configurations; it is
  strictly correct if the (anisotropic) hyperfine interaction stems
  predominantly from coupling to a single nuclear spin.}
\end{thebibliography}

\end{document}